\documentclass{sig-alt-full}

\usepackage{amssymb,amsmath,graphicx}
\usepackage{algorithm}
\usepackage{algorithmic}

\newtheorem{theorem}{Theorem}[section]
\newtheorem{lemma}[theorem]{Lemma}
\newtheorem{claim}[theorem]{Claim}

\newtheorem{proposition}[theorem]{Proposition}

\newtheorem{definition}{Definition}[section]

\newtheorem{remark}{Remark}[section]

\newenvironment{nestproof}[1]{\noindent{\it Proof of #1:
  }}{\qed\par}
\newenvironment{proof-sketch}{\par \noindent {\it Sketch of Proof.}}{\qed \par}

\newcommand{\op}{\sf Opt}
\newcommand{\ceil}[1]{\lceil #1 \rceil}
\newcommand{\floor}[1]{\lfloor #1 \rfloor}

\newcommand{\junk}[1]{}

\begin{document}

\newif\iffull
\fulltrue

\newif\ifproc
\procfalse

\conferenceinfo{SPAA'07,} {June 9--11, 2007, San Diego, California, USA.}
\CopyrightYear{2007}
\crdata{978-1-59593-667-7/07/0006} 


\title{Approximation Algorithms for Multiprocessor Scheduling under
Uncertainty}
\ifproc
\subtitle{[Extended Abstract]}
\fi

\numberofauthors{2} 
\author{
\alignauthor
Guolong Lin\titlenote{Part of this work was done when the author was
at Northeastern University.} \\
   \affaddr{Akamai Technologies} \\
   \affaddr{8 Cambridge Center, Cambridge, MA 02142} \\
   \email{glin@akamai.com} 
\alignauthor
Rajmohan Rajaraman \\
   \affaddr{College of Computer and Information Science} \\
   \affaddr{Northeastern University, Boston MA 02115} \\
   \email{rraj@ccs.neu.edu}
   }

\maketitle


\begin{abstract} Motivated by applications in grid computing and
project management, we study multiprocessor scheduling in scenarios
where there is uncertainty in the successful execution of jobs when
assigned to processors.  We consider the problem of {\em
multiprocessor scheduling under uncertainty}, in which we are given
$n$ unit-time jobs and $m$ machines, a directed acyclic graph $C$
giving the dependencies among the jobs, and for every job $j$ and
machine $i$, the probability $p_{ij}$ of the successful completion of
job $j$ when scheduled on machine $i$ in any given particular step.
The goal of the problem is to find a schedule that minimizes the
expected makespan, that is, the expected completion time of all the
jobs.

The problem of multiprocessor scheduling under uncertainty was
introduced by Malewicz and was shown to be NP-hard even when all the
jobs are independent.  In this paper, we present polynomial-time
approximation algorithms for the problem, for special cases of the dag
$C$.  We obtain an $O(\log n)$-approximation for the case of
independent jobs, an $O(\log m \log n
\log(n+m)/\log\log(n+m))$-approximation when $C$ is a collection of
disjoint chains, an $O(\log m \log^2 n)$-approximation when $C$ is a
collection of directed out- or in-trees, and an $O(\log m \log^2 n
\log(n+m)/\log\log(n+m))$-approximation when $C$ is a directed forest.
\end{abstract}




\newcommand{\suu}{{\mbox{\sf SUU}}}
\newcommand{\suui}{{\mbox{\sf SUU-I}}}
\newcommand{\suuc}{{\mbox{\sf SUU-C}}}
\newcommand{\suut}{{\mbox{\sf SUU-T}}}
\newcommand{\msp}{\sf MaxSumProb}
\newcommand{\msm}{\sf MaxSumMass}
\newcommand{\msme}{\sf MaxSumMass-Ext}
\newcommand{\algmsm}{\mbox{\sc MSM-ALG}}
\newcommand{\algmsme}{\mbox{\sc MSM-E-ALG}}
\newcommand{\algsuui}{\mbox{\sc SUU-I-ALG}}
\newcommand{\algsuuiObl}{\mbox{\sc SUU-I-OBL}}

\newcommand{\topt}{T^{\mbox{\tiny OPT}}}
\newcommand{\oblivs}{oblivious schedule}
\newcommand{\pseus}{pseudo-schedule}
\newcommand{\concat}[2]{#1\circ #2}


\section{Introduction}
We study the problem of multiprocessor scheduling under uncertainty,
which was introduced in~\cite{Malewicz-spaa05} to study scenarios
where there is uncertainty in the successful completion of a job when
assigned to a server.  One motivating application is in grid
computing, where a large collection of computers, often geographically
distributed,cooperate to solve complex computational tasks. To make
better use of the distributed computers, a task is usually divided
into smaller pieces (or jobs) and handed to different computers.  For
many applications, there could be non-trivial dependencies among these
jobs.  Due to the possible physical failures, or simply the
distributed nature of the computing environment, a machine may not
successfully execute the assigned job on time.  In this scenario, a
natural goal is to determine a schedule of assigning the given jobs to
the computers so that the expected completion time of the task is
minimized.

A similar example, also discussed in~\cite{Malewicz-spaa05}, arises
while managing a large project in an organization.  The project may be
broken down into small jobs with dependencies among them, i.e., a job
may be executed only after the successful completion of another set of
jobs.  A group of workers are assigned to this project.  Due to
practical reasons and different skills, a worker may not be able to
finish an assigned job successfully on time.  To decrease the chance
of the potential delay of some key jobs, the project manager could
(and would want to) assign several workers to these jobs at the same
time.  Based on past experiences and the workers' skill levels, the
project manager can estimate the successful probability of any
particular worker finishing any particular job. The challenge for the
manager is to work out a strategy (or schedule) of assigning the
workers to the jobs so that the {\em expected} completion time of the
whole project is as small as possible.

Motivated by the examples above, we study the problem of {\em
  multiprocessor scheduling under uncertainty}, henceforth referred to
as {\suu}.  We have a set of $m$ machines, a set of $n$ unit-time
jobs, and a directed acyclic graph representing precedence constraints
on the order of the execution of the jobs.  We are also given, for
every job $j$ and machine $i$, the probability $p_{ij}$ of the
successful completion of job $j$ when scheduled on machine $i$ in any
given particular step.  To compensate for this uncertainty, multiple
machines can be assigned to one job at the same time.  We focus on the
problem of computing a schedule to minimize the expected time to
complete all the jobs, i.e., the expected makespan.


\subsection{Our results} \label{sec:sch.work}
The multiprocessor scheduling problem {\suu} is shown to be NP-hard
in~\cite{Malewicz-spaa05} even when all jobs are independent.  In this
paper, we present approximation algorithms for \suu, for several
special classes of dependency graphs.
\begin{itemize}
\item 
We first consider the case when all the jobs are independent and
present an $O(\log n)$-approximation algorithm for the problem
(\S\ref{sec:sch.ind}).
\junk{We later
improve the preceding bound to $O(\log n\cdot\log(\min\{n, m\}))$
based on a linear programming approach (\S\ref{sec:sch.ind.improved}).}
\end{itemize}
A crucial component of our approach to the independent jobs case is
the formulation of a sub-problem in which we aim to maximize the sum
of success probabilities for the jobs.  A similar strategy, refined to
handle job dependencies, allows us to attack the more general case
where the jobs are not independent.
\begin{itemize}
\item When the precedence constraints on the jobs form a collection of
  disjoint chains, we obtain an \\ $O(\log m\log n\frac{\log
    (n+m)}{\log\log (n+m)})$ approximation algorithm in
  (\S\ref{sec:sch.chain}). Our results rely on solving a (relaxed)
  linear program and rounding the fractional solution using results
  from network flow theory.
\item Using the algorithm for disjoint chains and the chain
    decomposition techniques of~\cite{KumarEtAl05}, we obtain \\ $O(\log
  m \log^2 n)$ and $O(\log m\log^2 n\frac{\log (n+m)}{\log\log
  (n+m)})$ approximations for a collection of in- or out-trees and
  directed forests, respectively (\S\ref{sec:sch.forest}).
\end{itemize}
The schedules computed by the algorithms for disjoint chains,
trees, and directed forests, are all oblivious in the sense that
they specify in advance the assignment of machines to jobs in each
time step, independent of the set of unfinished jobs at that step.
Oblivious schedules are formally defined in \S\ref{sec:sch.def}, where
we also present useful definitions and important properties of
schedules that are used in our main results.

To the best of our knowledge, our results are the first approximation
algorithms for multiprocessor scheduling under uncertainty
problems.\junk{ Malewicz~\cite{Malewicz-spaa05} gives a
polynomial-time (exact) algorithm to the problem for the special case
where the {\em width} of the dependency graph, i.e, the maximum number
of independent jobs, and the number of machines are both bounded.
We believe, however, that the approximation factors we have obtained
can be improved.  We feel that our general approach presented
here can be extended to attack the even more general case where the
dependency graph is a general directed acyclic graph (dag).  We
discuss open problems in \S\ref{sec:sch.open}.  }
\junk{Due to space constraints, we have omitted many of the proofs; they may
be found in the appendices~\ref{app:def} through~\ref{app:dep}.}

\junk{This portion of the paper is based on joint work with Rajmohan
Rajaraman.}


\subsection{Related work} \label{sec:sch.related}

The problem studied in our work was first defined in the recent work
by Malewicz~\cite{Malewicz-spaa05}, largely motivated by the
application of scheduling complex dags in grid
computing~\cite{Foster99}.\junk{where, as usual, we have $m$ machines
to finish $n$ unit-time jobs observng a dag-like precedence
constraints. The major difference is that when machine $i$ is assigned
to job $j$, there is a probability of success $p_{ij}$, which may not
be $1$. One motivation for the problem is in grid
computing~\cite{Foster99}, where a computer may fail to correctly
execute an assigned job, which could delay the whole process of
execution of the system.  Another motivation comes from the project
management or production planning. The workers are assigned to a
collection of tasks. Due to factors such as different skill levels,
the workers may not be able to finish the assigned tasks within the
deadlines. This uncertainty of finishing the tasks are modeled by the
(unrelated) probabilities values $p_{ij}$'s.  } Malewicz characterizes
the complexity of the problem in terms of the number of the machines
and the {\em width} of the dependency graph, which is defined as the
maximum number of independent jobs. He shows that when the number of
machines and the width are both constants, the optimal regimen can be
computed in polynomial time using dynamic programming. However, if
either parameter is unbounded, the problem is NP-hard. Also, the
problem can not be approximated within a factor of $5/4$ unless
P=NP. Our work extends that of Malewicz by studying the
approximability of the problem when neither the width of the dag nor
the number of machines is bounded.

The uncertainty of the scheduling problem we study comes from the
possible failure by a machine assigned to a job, as modeled by the
$p_{ij}$'s.  There have been different models of uncertainty in the
scheduling literature. Most notable is the model where each task has a
duration of random length and may require different amount of
resources. For related work, see
\cite{Fernandez-98J,Fernandez-98C,Herroelen-05J,Skutella-soda01,KleRabTar00,Goel-focs99}.


Scheduling in general has a rich history and a vast
literature. \junk{One familiar example is the scheduling of classrooms
for classes or exams.The typical setup for the scheduling problem is
to complete $n$ tasks (or jobs) using $m$ machines (or processors).}
There are many variants of scheduling problems, depending on various
factors. For example: Are the machines related? Is the execution
preemptive? Are there precedence constraints on the execution of the
jobs? Are there release dates associated with the jobs? What is the
objective function: makespan, weighted completion time, weighted flow
time, etc.? See~\cite{Hall97} for a survey and
~\cite{Graham66,LenShmTar90,Skutella-jacm01, LeightonMR94, ChuShm99,
KumarEtAl05} for representative work.

Two particular variants of scheduling closely related to our work is
  job shop scheduling~\cite{ShmSteWei94} and the scheduling of
  unrelated machines under precendence constraints. In the job shop
  scheduling problem, we are given $m$ machines and $n$ jobs, each job
  consisting of a sequence of operations. Each operation must be
  processed on a specified machine.  A job is executed by processing
  its operations according to the associated sequence. At most one job
  can be scheduled on any machine at any time. The goal of the job
  shop scheduling problem is to find a schedule of the jobs on the
  machines that minimizes the maximum completion time. This problem is
  strongly NP-hard and widely
  studied~\cite{GareyJohnson,LawlerEtAl91,ApplegateCook}.\junk{, and
  is very difficult in practical computation. Applegate and
  Cook~\cite{ApplegateCook} have posed several benchmark instances as
  open problems. These job shop instances have only $15$ jobs, $15$
  machines and $225$ operations and yet have not been solved
  efficiently by most known methods.} Also extensively studied is
  the problem of preemptively scheduling jobs with precedence
  constraints on unrelated parallel
  machines~\cite{LeightonMR94,ShmSteWei94,KumarEtAl05}, the processing
  time of a job depends on the machine to which it is assigned.  One
  common characteristic of this problem and {\suu} is that in each
  problem, the capability of a machine $i$ to complete a job $j$ may
  vary with both $i$ and $j$.  However, while the unrelated parallel
  machines problem models this nonuniformity using deterministic
  processing times that vary with $i$ and $j$, in {\suu} the jobs are
  all unit-size but may fail to complete with probabilities that vary
  with $i$ and $j$.  Owing to the uncertainty in the completion of
  jobs, {\suu} schedules appear to be more difficult to specify and
  analyze.  One other technical difference is that in {\suu} we allow
  multiple machines to be assigned to the same job at the same time,
  for the purpose of raising the probability of successfully
  completing the job.  The unrelated parallel machines problem is
  typically solved by a reduction to instances of the job shop
  scheduling problem.  Some of our {\suu} algorithms also include
  similar reductions.

\newcommand{\jb}[1]{\mbox{Jobs}(#1)}
\section{Schedules, success \\probabilities, and mass} 
\label{sec:sch.def} 
In this section, we present formal definitions of a schedule
(\S~\ref{sec:sch.def.schedule}), introduce the notion of the mass of a
job and prove a key technical theorem about the accumulation of mass
of a job within the expected makespan of a given schedule
(\S~\ref{sec:sch.def.mass}).
\subsection{Schedules}
\label{sec:sch.def.schedule}
In {\suu}, we are given a set $J$ of $n$ unit-step jobs, and a set $M$
of $m$ machines. There are precedence constraints among the jobs,
which form a directed acyclic graph (dag) $C$. A job $j$ is {\em
eligible} for execution at step $t$ if all the jobs preceding $j$
according to the precedence constraints have been successfully
completed before $t$.  For every job $j$ and machine $i$, we are also
given $p_{ij}$, which is the probability that job $j$ when scheduled
on a machine $i$ will be successfully completed, {\em independent} of
the outcome of any other execution.  Multiple machines can be assigned
to the same job at the same step.  Without loss of generality, we
assume that for each $j$, there exists a machine $i$ such that $p_{ij}
> 0$.
\junk{
We first informally motivate our definition of a {\em schedule}, which
will be provided below.  Notice that our goal is to assign the machines
to the jobs at every step so as to complete all the jobs quickly, in
the {\em expected} sense. Such machine assignments may not be simple,
however, due to the execution uncertainties coming from the
$p_{ij}$'s. For example, if all the $p_{ij}$'s are less than $1$,
given any machine assignment for step $1$, there could be many
different execution outcomes (the set of unfinished jobs) at the end of step
$1$. In other words, for step $2$, one may need many different machine
assignment functions, depending on the possible outcomes from step
$1$. This is also true for steps beyond $2$.
}
\begin{definition}
  A {\bf schedule} $\Sigma$ of {\bf length} $T\in \mathbb{Z^+}\cup
  \{\infty\}$ is a collection of functions $\{f_{S,t}: M\to
  J\cup\{\perp\}\, |\, S\subseteq J,\; 1\le t < T+1\}$\junk{~\footnote{In
    this paper, we adopt the convention that $\infty+k =\infty$ for
    any finite $k$.}}. An {\bf execution} of the schedule $\Sigma$
  means that, at the start of each step $t$, if $S$ is the set of
  unfinished jobs: machine $i$ is assigned to job $f_{S,t}(i)$ if
  $f_{S,t}(i)$ is eligible and belongs to $S$; otherwise, $i$ is idle
  for that step.
\end{definition}
\junk{
{\bf Problem (Scheduling under Uncertainty (\suu)):} Given the job set
$J$, machine set $M$, precedence constraints $C$, and the $p_{ij}$'s,
as described earlier, the goal of the problem is to find a schedule
that minimizes the expected makespan, where makespan is the maximum
completion time of the jobs.
}
\junk{
If the maximum value of $t$ for which there is some $f_{S,t}$
specified is {\em finite}, we define this maximum value to be the
schedule's {\bf length}. Otherwise, we say that the schedule's length
is {\em infinite}. 
}

\junk{
We now formally define the notion of schedule used in this paper, as
well as a related notion of regimen, which will be followed by some
remarks.
}
\junk{
A schedule is {\em feasible} if at any time, each machine is assigned
to at most one job. A job $j$ is {\em eligible} for execution at some
time $t$ if all the jobs preceding $j$ according to the precedence
constraints have been successfully completed before $t$. If at some
time $t$ a job $j$ is not eligible while there are machines assigned
to $j$, the execution on $j$ will be a {\em failure} operation by
default, regardless of the values of the $p_{ij}$'s.
}

Our formal definition of a schedule specifies assignment functions
$f_{S,t}$ for infinite $t$. This is because there is a positive
probability for a job $j$ to be not completed yet by any given step if
$\forall i, p_{ij} < 1$.  For the purposes of optimizing expected
makespan, however, we can restrict our attention to a restricted class
of schedules.

\begin{definition}[\cite{Malewicz-spaa05}]
  A {\bf regimen} $\Sigma_g$ is a schedule in which $f_{S,t_1}(\cdot)
  = f_{S, t_2}(\cdot)$ for any $S\subseteq J$ and $t_1\neq t_2$. In
  other words, the assignment functions $f_{S,t}$'s depend only on the
  unfinished job set $S$. Thus, we can specify $\Sigma_g$ by a
  complete collection of functions $\{f_S: M\to S\cup\{\perp\} \,|\,
  S\subseteq J\}$. \junk{To use $\Sigma^g$, at the start of each time
    step, if $S$ is the set of unfinished jobs, $f_S$ will be used to
    assign the machines to the jobs for that step.}
\end{definition}
\junk{
At any step $t$, one may also need many
$f_{S,t}$'s because the unfinished job set $S$ at step $t$ is a {\em
random variable} due to the execution uncertainty.
}

We denote the minimum expected makespan for a given {\suu} instance by
{$\topt$}, which is finite because for any job $j$, there exists a
machine $i$, such that $p_{ij} > 0$.  It is not hard to see that there
exists an optimal schedule which is a regimen because at any step $t$,
one can determine an optimal assignment function, which only depends
on the subset of unfinished jobs at step $t$ and is independent of the
past execution history or the value $t$.  While a naive specification
of an arbitrary regimen uses $2^n$ different assignment functions,
certain regimens can be specified succinctly, for instance, by a
polynomial-length function that takes $S$ as input and returns $f_S$.
In this paper, we also consider a different restricted class of
schedules, called {\em oblivious schedules}.
\junk{The {\oblivs}s
we compute in this paper can be specified in space polynomial in the
size of the input.}

\begin{definition}
  An {\bf \oblivs} is a schedule in which every assignment function
  $f_{S,t}$ is independent of $S$, i.e., for all $t,S,S'$,
  $f_{S,t}(\cdot) = f_{S',t}(\cdot)$. Hence, the assignment functions
  at any step $t$ can be specified by a single function, which we
  denote by $f_t$.
\end{definition}

\junk{
\begin{remark} \label{rem:oblivious} If at some $t' > T$, there is
  some job still unfinished, then to use the {\oblivs} $\Sigma^o$, we
  need to specify an assignment function $f_{t'}(\cdot)$. In many
  situations, we simply {\em map} $t'$ to a $t\in [1,T]$, and use
  $f_t$ of $\Sigma^o$ as the assignment function for time $t'$, i.e.,
  $f_{t'}(\cdot) = f_t(\cdot)$. The specification of $f_{t'}(\cdot)$,
  or the mapping from $t'$ to $t$, will be context-dependent. Although
  sometimes we do not {\em explicitly} mention the assignment
  functions $f_{t'}$ for $t' > T$ when we discuss an {\oblivs}, we
  will specify unambiguously such functions whenever we analyze the
  expected makespan of the {\oblivs}.
\end{remark}
}

Oblivious schedules are appealing for two reasons. First, at any step
$t$, only one assignment function is needed, regardless of the actual
unfinished job set $S$ occurring at step $t$. Recall that there could
be many different such $S$ at a given $t$ because of the execution
uncertainty. The second benefit is more technical: {\oblivs}s allow
us to address the {\em uncertainty} in the {\suu} problem by solving
related {\em deterministic} optimization problems.

\subsection{Success probabilities and mass}
\label{sec:sch.def.mass}
When a subset of machines $S\subseteq M$ is assigned to $j$ in any
time step, the probability that $j$ is successfully completed is
$1-\prod_{i\in S} (1-p_{ij})$.  For ease of approximation, the
following Proposition is useful to us.
\begin{proposition} \label{pro:bounds}
Given $x_1, \cdots, x_k \in [0,1]$, $1-(1-x_1)\cdots (1-x_k) \le
x_1+\cdots + x_k$. Furthermore, if $x_1+\cdots + x_k \le 1$, then
$1-(1-x_1)\cdots (1-x_k) \ge e^{-1} (x_1+\cdots +x_k)$.
\end{proposition}

\begin{proof}
  The first assertion follows from the identity $(1-x_1)\cdots
  (1-x_k)\ge 1-(x_1+\cdots + x_k)$, which can be proved using a simple
  induction argument. The base case of $k=1$ is trivial. Suppose the
  identity holds for $k-1$. If $x_1+\cdots+x_{k-1} > 1$, then the
  identity holds for $k$; Otherwise, according to the induction
  hypothesis,
  \begin{eqnarray*}
  & & (1-x_1)\cdots (1-x_{k-1})(1-x_k)\\
  &\ge&  [1-(x_1+\cdots+x_{k-1})](1-x_k)\\
  &\ge &1-(x_1+\cdots+x_{k}). 
  \end{eqnarray*}
  For the second assertion, notice that if $0\le x\le 1$, $1-x \le
  e^{-x} \le 1-\frac{x}{e}$. Since $1-x \le e^{-x}$, $(1-x_1)\cdots
  (1-x_k) \le e^{-x_1}\cdots e^{-x_k}$, we have
  \begin{eqnarray*}
     & & 1-(1-x_1)\cdots (1-x_k) \\
     & \ge & 1- e^{-x_1}\cdots e^{-x_k} \\
     & = & 1-e^{-(x_1+\cdots+x_k)} \\
     & \ge & \frac{x_1+\cdots +x_k}{e},
  \end{eqnarray*}
  where the last inequality follows because $e^{-x} \le 1-\frac{x}{e}$
  for $x\in [0,1]$ and the assumption that $x_1+\cdots + x_k \le 1$.
\end{proof}

Proposition~\ref{pro:bounds} suggests that we can approximate the
success probability with a convenient linear form. \junk{For simplicity
of presentation, we introduce a name for this form.  In
  preparation for the presentation below, we introduce several new
  notions. We say a schedule is {\em oblivious} if at each step
  $t$, the scheduling function $f_{S,t}$ is independent of $S$.  }

\begin{definition}
For any schedule $\Sigma$, we define the {\bf mass} of a job $j$ at
  the end of step $t$ to be the sum, over all time $t' \in [1,t]$ and
  over every machine $i$ to which $j$ is assigned at time $t'$, of
  $p_{ij}$.  Thus, for an arbitrary schedule, the mass of a job $j$ at
  time $t$ is a random variable.  For an {\em \oblivs} $\Sigma_o$, the
  mass of $j$ at the end of any step $t$ is simply $$\min\{\sum_{1\le
  \tau\le t} \sum_{i:f_{\tau}(i) = j} p_{ij}, 1\},$$ where
  $f_{\tau}(\cdot)$ is the assignment function of $\Sigma_o$ at step
  $\tau$. We say that $j$ {\em accumulates} that mass by step $t$.
\end{definition}
\junk{Finally, an
  oblivious schedule is {\em nice} if every job accumulates at least
  constant mass by some time $t$.

Note that our goal is to use mass to approximate success probability.
We hence cap the value of the mass by $1$. See also Problem {\msm} in
\S\ref{sec:sch.ind.max} for more discussion.
}

The following theorem is crucial for our approach to the scheduling
problem. We emphasize that it holds for an arbitrary {\suu} instance.
It is used in the proofs of Theorem~\ref{thm:exist} and
Lemma~\ref{lem:link}.

\begin{theorem}\label{thm:sim}
  Let $\Sigma$ be a schedule for an {\suu} instance, whose expected
  makespan is $T$. For any job $j$, in an execution of $\Sigma$ for
  $2T$ steps, with probability at least $1/4$, $j$ accumulates a mass
  of at least $1/4$.
\end{theorem}
\begin{proof}
  Let $A$ be the event that $j$ is finished within step $2T$. Let
  $S_t$ be the random variable denoting the collection of machines
  assigned to job $j$ at step $t$ and $P(S_t) = \sum_{i\in S_t}
  p_{ij}$. Let $B$ be the event that $\sum_{1\le t\le 2T}{P(S_t)} \le
  1/4$. What we want to prove is $\Pr(B^c) \ge 1/4$. Observe that
  $\Pr(A)$ equals $\Pr(A\cap B) + \Pr(A\cap B^c)$, which is at most
  $\Pr(A\cap B) + \Pr(B^c)$.

We estimate the value of $\Pr(A\cap B)$ below. Observe that all
possible executions of $\Sigma$ on the jobs form an infinite rooted
tree, in which each node represents an intermediate state during an
execution (see Figure~\ref{fig:schedule} for an illustration). Each
node has an associated set of jobs, representing the unfinished jobs
at that state. For a node $N$, let $\jb{N}$ be its associated set of
unfinished jobs. Note that $\jb{R}$ for the root node $R$ at level $0$
consists of the entire set of jobs. The nodes at level $k$ denote the
states after $k$ steps. From each node $N$ at level $k$ to each node
$Q$ at level $k+1$, we can compute the corresponding transition
probability according to the assignment function $f_{\tiny
\jb{N},k+1}$.  \junk{We prove the technical lemma below.}

\begin{figure}[h]
\centering
\resizebox{\columnwidth}{!}{\includegraphics{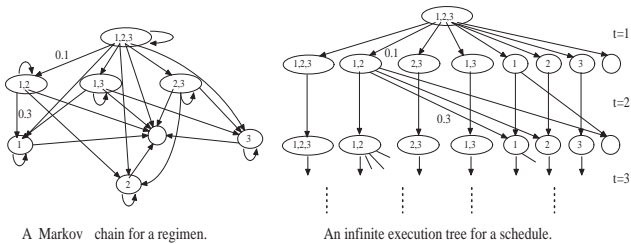}}
\caption{\small An illustration of the schedule. For simplicity purpose, we
  only use $3$ jobs. Each node represents an intermediate state, with
  its associated set of unfinished jobs appearing inside. The number
  close to an edge represents its transition probability. The left
  graph is a Markov chain representation of a regimen. The right graph
  is a rooted tree representation of the execution of a schedule. To
  avoid cluttering, we only show the complete transitions for nodes
  $\{1,2\}$ and $\{1\}$ at step $2$.}
\label{fig:schedule}
\end{figure}

\begin{lemma}
  Consider a tree node $N$ at level $k$, where $j\in \jb{N}$. For
  $1\le t\le k$, let $S_t$ be the machine set assigned to $j$ during
  step $t$ along the path leading to $N$ from $R$. Assume that
  $\sum_{1\le t\le k}P(S_t) \le c$, where $c\le 1$. And let $P(j,N)$
  be the probability that $j$ will be finished by level (step) $2T$
  following a tree path through $N$ and $\sum_{1\le t\le 2T}P(S_t) \le
  c$. Then $P(j,N) \le c - \sum_{1\le t\le k} P(S_t)$.
\end{lemma}

\begin{nestproof}{Lemma}
  We prove the lemma by backward induction on the level number $k$.
  Consider the base case: $N$'s level is $2T-1$. We only need to
  execute the schedule for one more step. Let $S_{2T}$ be the set
  of machines assigned to $j$ during step $2T$. If $P(S_{2T}) >
  c-\sum_{1\le t\le 2T-1} P(S_t)$, then $P(j,N) = 0$. Otherwise, the
  probability that $j$ is finished within this step is at most
  $P(S_{2T})$. In either case, the claim is true. 

  We now assume that the claim is true for any level $k\le 2T-1$, our
  aim is to prove that the claim is also true for level $k-1$.
  Consider a tree node $N$ at level $k-1$. Let $S_k$ be the set of
  machines assigned to $j$ during step $k$ according to assignment function
  $f_{\tiny \jb{N},k}$. A child node of $N$ at level $k$ either does
  not contain $j$ ($j$ is finished at step $k$) or contains $j$ ($j$
  is not finished at step $k$). Let the probabilities of the two cases
  be $P_1$ and $1-P_1$, respectively. Denote all the children nodes
  where $j$ is still unfinished as $L$.

If $P(S_k) > c-\sum_{1\le t\le k-1} P(S_t)$, then $P(j,N) = 0$,
which is $\le c-\sum_{1\le t\le m-1} P(S_t)$.
Otherwise,
\begin{eqnarray*}
  P(j,N) & = &P_1 + \sum_{Q\in L} P(j,Q) \\
  &\le & P_1 + \sum_{Q\in L} (c- \sum_{1\le t\le k} P(S_t)) \\
  & = & P_1 + (1-P_1) (c- \sum_{1\le t\le k} P(S_t)) \\
  & \le & P_1 + (c- \sum_{1\le t\le k} P(S_t)) \\
  &\le & c- \sum_{1\le t\le k-1} \Pr(S_t),
\end{eqnarray*}
where the second inequality follows from the induction hypothesis and
the last inequality follows from the fact that $P_1 \le P(S_k)$.  This
proves the induction step and hence the Lemma.
\end{nestproof}

By invoking the lemma with $c=1/4$, we obtain $\Pr(A\cap B) = P(j, R)
\le c = 1/4$. Hence $\Pr(A) \le 1/4 + \Pr(B^c)$.  And by Markov's
inequality, $\Pr(A) \ge 1/2$. We conclude that 
$\Pr(B^c) \ge 1/4$, completing the proof.
\end{proof}

\junk{
We have thus proved the following claim (even when the jobs have
precedence constraints):
\begin{claim} \label{clm:sim}
  If we simulate the regimen $\Sigma$ for $2T$ steps, for any job $j$,
  with a chance of at least $1/4$, $j$ ``follows'' a path with total
  success probabilities~\footnote{We approximate total success
    probability by $\sum P(S_t)$, see the discussion in
    \S\ref{sec:sch.modprob}.} at least $1/4$.
\end{claim}
} 

\section{Independent jobs} \label{sec:sch.ind}
In this section, we study a special case of the scheduling problem,
where the jobs are independent. We refer to this problem as {\suui}.
\junk{{\bf Problem (Scheduling under Uncertainty - Independent (\suui)):} An
instance of \suu, where the dependency graph $C$ has no edges, i.e.,
the jobs $J$ are independent.}  To compute a solution to {\suui}, we
first establish that there exists an {\oblivs} in which the total mass
accumulated by the jobs in $O(\topt)$ steps is $\Omega(n)$.  To find
such a schedule, we formulate a subproblem for maximizing the total
sum of masses and then give polynomial-time algorithms to compute an
$O(\log n)$-approximate schedule and an $O(\log^2 n)$-approximate
oblivious schedule for {\suui}.  For oblivious schedules, we improve
the approximation factor to $O(\log n\cdot
\log(\min\{n,m\}))$ when we study the more general case with
chain-like precedence constraints in \S\ref{sec:sch.chain}.  \junk{The
reason for presenting the weaker bound for the independent jobs is
two-fold. First, The approach for the independent jobs provides key
ideas for solving the more general case. Second, the weaker bound in
this section is based a {\em simpler} combinatorial algorithm.}


\begin{theorem} \label{thm:exist} If
  there exists a schedule $\Sigma$ for {\suui} with expected makespan
  $T$, then there exists an {\oblivs} of length $2T$, in which the
  total mass accumulated by all jobs is at least $n/16$.
\end{theorem}
\begin{proof}
Consider an execution $E$ of $\Sigma$ for $2T$ steps. This execution
yields naturally an {\oblivs} $\Sigma_E$ of length $2T$, whose
assignment functions $f_t(\cdot)$'s are defined as follows: $f_t(i) =
j$ if machine $i$ is assigned to job $j$ at step $t$ in $E$. Note that
due to execution uncertainty, $E$, and hence $\Sigma_E$ are both {\em
random variables}. By Theorem~\ref{thm:sim}, for any job $j$, with
probability at least $1/4$, $j$ accumulates a mass of at least $1/4$
by step $2T$ in $\Sigma_E$.  Thus, the expected mass of $j$ at step
$2T$ in $\Sigma_E$ is at least $1/16$.  This implies that the expected
total mass of all the jobs at step $2T$ in $\Sigma_E$ is at least
$n/16$.  Therefore, there exists an oblivious schedule in which the
total mass of the jobs at step $2T$ is at least $n/16$.
\end{proof}
\iffull
\subsection{An $O(\log n)$-approximate schedule for {\suui}}
\label{sec:sch.ind.approx}
\fi

\junk{We are now in a position to show how to compute an oblivious
  schedule whose existence is proved in Theorem~\ref{thm:exist}. } 

Motivated by Theorem~\ref{thm:exist}, we formulate subproblem {\msm}
for maximizing the sum of masses.  In {\msm}, we are given a set $J$
of $n$ independent, unit-step jobs, a set $M$ of $m$ machines, and the
probabilities $p_{ij}$, and the goal is to find an assignment $f: M\to
J\cup\{\perp\}$ for a single step that maximizes the sum of masses
over the jobs in the step.  In Figure~\ref{fig:ind}, we present a
$1/3$-approximation algorithm {\algmsm} for {\msm} (which can be shown
to be NP-hard), and our approximation algorithm for {\suui}, which
simply executes, in every step, {\algmsm} on the unfinished jobs.
\junk{Algorithm
{\algmsm} will be used in \S\ref{sec:sch.ind.approx}, after some
slight extension, to compute an {\oblivs} for the {\suui} problem,
with an expected makespan of $O(\log^2 n)\cdot\topt$.}

\begin{figure*}[bt]
\begin{center}
\fbox{
{\small
\begin{minipage}[bt]{3in}
{\bf Algorithm} {\algmsm}\\
{\bf INPUT:} Jobs $J$, machines $M$, $p_{ij}$'s.
\begin{list}{\labelitemi}{\setlength{\leftmargin}{0.1in}\setlength{\itemsep}{0ex}}
\item Set $f(i)$ to nil, $i\in M$.
\item For each $p_{ij}$ in nonincreasing order: If $f(i)$ is nil and
$\sum_{x:f(x)= j} p_{xj} + p_{ij} \le 1$, assign $i$ to $j$, i.e.,
    $f(i) \gets j$. 
\item For every unused machine $i$, $f(i) \gets \perp$; output $f$.
\end{list}
\end{minipage}
\hspace*{0.2in}
\begin{minipage}[bt]{3in}
{\bf Algorithm} {\algsuui}\\
{\bf INPUT:} Jobs $J$, machines $M$, $p_{ij}$'s.
\begin{list}{\labelitemi}{\setlength{\leftmargin}{0.1in}\setlength{\itemsep}{0ex}}
\item Let $S_t$ denote the set of unfinished jobs at the start of step $t$
\item 
In each step $t$, schedule according to the assignment determined by
{\algmsm} applied to $S_t$ and all machines.
\end{list}
\end{minipage}
}
}
\end{center}
\caption{An approximation algorithm for scheduling independent jobs.
\label{fig:ind}}
\end{figure*}

\junk{
    \item If $f(i) \ne nil$ (some previous $p_{ij'}$ is used), throw
      away $p_{ij}$. Next.
    \item If $\sum_{x:f(x)= j} p_{xj} + p_{ij} > 1$, throw away
      $p_{ij}$. Next.
    \item Otherwise, assign $p_{ij}$ to $j$, i.e., $f(i) \gets j$.
}
\begin{theorem} \label{thm:msm} {\algmsm} computes a
      $1/3$-approximate solution to Problem {\msm}. \qed
\end{theorem}

\begin{proof}
  Consider a bi-partite graph, where one side of the graph lie the
  nodes for jobs $J$ and the other side lie the nodes for machines
  $M$. There is an edge $(i,j)$ between machine $i$ and job $j$ for
  any $p_{ij} > 0$. {\algmsm} can be viewed as picking and orienting
  the edges. Let Opt = \{$(i,j)$\} be the collection of edges of
  picked by the optimum assignment $f^*$. Let {\sc Sol} be the
  solution computed by {\algmsm}. We use a charging argument below.
  Consider any edge $(i,j)\in$ Opt.
\begin{enumerate}
\item $(i,j) \in$ {\sc Sol}, charge $p_{ij}$ to itself.
\item $(i,j) \notin$ {\sc Sol}:
  \begin{enumerate}
  \item $(i,j)$ is not added because in step 2, $f(i) \ne nil$.
    Let $j'= f(i)$. Charge $p_{ij}$ to $p_{ij'}$ where $(i,j')\in$
    {\sc Sol}. Notice that $p_{ij} \le p_{ij'}$, and $p_{ij'}$ will be
    charged at most once due to this situation because each machine
    $i$ in Opt is used at most once.
  \item $(i,j)$ is not added because in step 2, $f(i) = nil$ yet
    $\sum_{x:f(x)= j} p_{xj} + p_{ij} > 1$. Since $p_{ij}$'s are
    processed in decreasing order, we conclude that in {\sc Sol},
    $\sum_{x: f(x) = j} p_{xj} \ge 1/2$. Charge $p_{ij}$ to\\ $2\sum_{x:
      f(x) = j} p_{xj}$.
  \end{enumerate}
\end{enumerate}
Observe that one copy of {\sc Sol} is sufficient to cover the charges
of types 1 and 2(a). Two copies of {\sc Sol} are sufficient to cover
the charges of type 2(b) because, by definition, the mass of any job
is at most $1$ in any assignment.  \junk{$\sum_{i: f^*(i) = j} p_{ij}
  \le 1$.}

We conclude that {\algmsm} computes a solution with an approximation
factor $1/3$.
\end{proof}

\junk{In this section, we give an approximation algorithm for Problem
{\suui} (Theorem~\ref{thm:sch.ind}). Our approach is to accumulate for
every job some {\em constant} mass using as few steps as possible, and
then replicate each step's machine assignment $O(\log n)$ times so
that all jobs are finished with high probability within a desired
makespan bound using the replicated {\oblivs}.  The main remaining
question for us now is how to compute the {\oblivs} of short length in
which every job accumulates a constant mass, which we now address.
}
\begin{theorem}
Algorithm {\algsuui} is an $O(\log n)$-approximation algorithm for
{\suui}.
\end{theorem}
\begin{proof}
Let $S_t$ denote the set of unfinished jobs at the start of step $t$.
Then, by Theorem~\ref{thm:exist}, there exists an {\oblivs} of length
$2\topt$ starting from step $t$, in which total mass of all jobs in
$S_t$ is at least $|S_t|/16$.  By averaging over the $2\topt$ time
steps of this schedule, there exists an assignment of jobs to machines
in step $t$ such that the total mass of the jobs in $S_t$ in step $t$
is at least $|S_t|/(32\topt)$.  By Theorem~\ref{thm:msm}, in step $t$
of {\algsuui}, the total mass of the jobs accumulated in step $t$ is
at least $|S_t|/(96\topt)$.  By Proposition~\ref{pro:bounds}, it
follows that the expected number of jobs that complete in step $t$ is
at least $|S_t|/(96e\topt)$.

We thus have a sequence of random variables $S_t$ which satisfy the
property $E[|S_{t+1}| \, | S_t] = |S_t|(1 - 1/(96e\topt))$.  By
straightforward Chernoff bound
arguments~\cite{chernoff:bound,hoeffding:bound}, we obtain that with
high probability, $S_t$ is empty within $O(\topt \log n)$ steps.
\end{proof}

\ifproc
The schedule computed by {\algsuui} is adaptive in the sense that the
assignment function for each step is dependent on the set of
unfinished jobs at the start of the step.  Using an extension of
{\algmsm}, we also obtain a polynomial-time combinatorial algorithm to
compute an {\em oblivious}\/ schedule with expected makespan within an
$O(\log^2 n)$ of the optimal.  Due to space constraints, we defer this
result to the full paper.  In \S\ref{sec:sch.chain}, we improve this
bound further to $O(\log n\cdot\log(\min\{n,m\}))$ using an LP-based
algorithm.
\fi

\iffull
\subsection{An approximate oblivious schedule for \suui}
\label{sec:ind.obl}
The schedule computed by {\algsuui} is adaptive in the sense that the
assignment function for each step is dependent on the set of
unfinished jobs at the start of the step.  Using an extension of
{\algmsm}, we develop in this section a polynomial-time combinatorial
algorithm to compute an {\em oblivious}\/ schedule with expected
makespan within an $O(\log^2 n)$ of the optimal.  In
\S\ref{sec:sch.chain}, we improve this bound further to $O(\log
n\cdot\log(\min\{n,m\}))$ using an LP-based algorithm.

According to Theorem~\ref{thm:exist}, there exists an {\oblivs} of
length $2\topt$, in which total mass of all jobs is at least
$n/16$. Intuitively, if one computes an oblivous schedule $\Sigma_1$
of length $2\topt$ with the aim of maximizing the {\em total sum} of
masses over the jobs, there should be {\em many} jobs accumulating
constant masses in $\Sigma_1$. One can then remove those jobs and
compute a second {\oblivs} $\Sigma_2$ of length $2\topt$ to maximize
the total sum of masses for the remaining jobs, to remove some
additional jobs which have accumulated constant masses. Since each
computation of the {\oblivs} removes {\em many} jobs, this process
should terminate quickly. By concatenating the $\Sigma_1,
\Sigma_2,\ldots$ together, one obtains an {\oblivs} $\Sigma$ in which
{\em every} job accumulates constant mass.

By Theorem~\ref{thm:msm}, we have a $1/3$ approximation algorithm for
Problem {\msm}.  However, {\msm} only considers {\oblivs}s of length
1, i.e., each machine is assigned to at most one job. What we need is
a procedure of finding an oblivous schedule of length $2\topt$, which
maximizes the sum of masses over jobs. It turns out that one can
extend {\algmsm} easily to take into account the schedule length,
which can be {\em arbitrary}, and still obtain the same aproximation
factor of $1/3$. We now formalize our discussion.

\junk{ Due to the discussion in
  \S\ref{sec:sch.modprob}, we approximate the success probabilities by
  the sum of those $p_{ij}$'s assigned to $j$, sacrificing only a
  constant factor. We call an oblivious schedule {\em nice} if every
  job accumulates at least $c_0$ (modified) success probability.}

{\bf Problem} ({\msme}): We are given a set $J$ of $n$ independent,
unit-step jobs and a set $M$ of $m$ machines. Let $p_{ij}$ denote the
probability that job $j$ is successfully completed if assigned to
machine $i$. We are also given a parameter $t\in \mathbb{Z^+}$. The
goal of the problem is to find an {\oblivs} $\Sigma_o$ of length $t$
such that the total sum of masses accumulated by the jobs by step $t$
is maximized.

  \junk{ (i) $\forall j, \sum_{1\le \tau\le t}
  \sum_{i:f_\tau(i) = j} p_{ij} \le 1$; (ii) $\sum_j \sum_{1\le
    \tau\le t} \sum_{i:f_\tau(i) = j} p_{ij}$ is maximized.  }

We show below Algorithm {\algmsme}, which outputs an {\oblivs}
$\Sigma_o$ of length $t\in \mathbb{Z^+}$ that is a $1/3$ approximate
solution to Problem {\msme}. Algorithm {\algmsme} is a simple
modification from {\algmsm} as follows. Since the schedule is of
length $t$, each machine can be assigned $t$ times. We maintain a {\em
  remaining capacity} parameter for each machine, $t_i$, initialized
to the value $t$, to keep track of how many steps machine $i$ is still
available to be assigned. We also use $x_{ij}$ to keep track of how
many steps machines $i$ is assigned to job $j$. In Step 2(a) of
{\algmsme}, as long as $t_i$ is positive, assign $i$ to $j$ for as
many steps as necessary. In Step 2(b), we update $t_i$ accordingly. In
Step 3, we output an {\oblivs} $\Sigma_o = \{f_{\tau}(\cdot): 1\le\tau\le
t\}$, which can be specified by $x_{ij}$'s as follows. Let
$j_1,\ldots, j_n$ be an ordering of the jobs. $f_{\tau}(i) = j_k$ for
$\sum_{1\le l< k} x_{ij_l} + 1\le \tau\le \sum_{1\le l\le k} x_{ij_l}$
and $1\le k\le n$. Observe that the running time of {\algmsme} is
independent of the value $t$ because each $p_{ij}$, hence each pair
$(i,j)$, is processed exactly once in Step 2. It is not hard to see
that {\algmsme} outputs a $1/3$ approximate solution to Problem
{\msme} because similar analysis for {\algmsm} from
Theorem~\ref{thm:msm} can be applied.

\begin{algorithm}
  \caption{\algmsme}
  \label{alg:msme}
  {\bf INPUT:} Jobs $J$, machines $M$, $p_{ij}$'s and $t$.
  \begin{enumerate}
  \item Sort $p_{ij}$'s in decreasing order. Initialize: $\forall i,
    t_i \gets t$; $\forall i, j, x_{ij} \gets 0$.
  \item For each $p_{ij}$ according to the order:
    \begin{enumerate}
    \item $x_{ij} \gets \min\left\{t_i, \left\lfloor\frac{1- \sum_{k\in M}
          x_{kj}\cdot p_{kj}}{p_{ij}}\right\rfloor\right\}$.
      \item $t_i \gets t_i - x_{ij}$.
    \end{enumerate}
  \item Output $\Sigma_o$ specified by $x_{ij}$'s.
  \end{enumerate}
\end{algorithm}

\begin{lemma} \label{lem:msme} {\algmsme} computes a solution to
  Problem {\msme} with an approximation factor $1/3$.
\end{lemma}

We now present an approximation algorithm {\algsuuiObl} for Problem
{\suui}.

\begin{algorithm} \label{alg:suui}
  \caption{\algsuuiObl}
  {\bf INPUT:} Jobs $J$, machines $M$, $p_{ij}$'s.
  \begin{enumerate}
  \item $t \gets 1$.
  \item $I \gets 1$. $R\gets J$. $\Sigma\gets$ ``empty schedule''.
  \item While ($|R| > 0$) and ($I\le 66\log n$)
    \begin{enumerate}
    \item Let $\Sigma_I$ be the output of invoking {\algmsme} on $R,M$
      with the current $t$ value. $\Sigma \gets
      \concat{\Sigma}{\Sigma_I}$.
    \item Remove jobs that accumulate at least $1/96$ mass from $R$.
    \item $I\gets I+1$.
    \end{enumerate}
  \item If $|R| > 0$, then $t\gets 2t$, GOTO step 2; Otherwise,
    return  $\Sigma$.
  \end{enumerate}
\end{algorithm}

A few comments on {\algsuuiObl} are in order.  We use {\algmsme}
repeatedly to accumulate constant masses for a good fraction of the
jobs each round, until all jobs accumulate constant masses. There is
still one obstacle though. Since we don't know the value of $\topt$,
we have to ``guess'' a value of $t$ for {\algmsme}, which must be
large enough, e.g., at least $2\topt$, to ensure that there {\em
exists} an {\oblivs} of length $t$ in which the total mass is at least
$n/16$, as proved in Theorem~\ref{thm:exist}. In summary, in the loop
of {\algsuuiObl} (Step 3), we repeatedly invoke {\algmsme} to
accumulate $1/96$ mass for the jobs, for at most $66\log n$ rounds (we
will explain the reason shortly). At the end of the loop (Step 4), if
there are some remaining jobs, that means our $t$ value is not large
enough, we hence double the value of $t$ and try the new $t$ again by
resetting the other parameters.  Note that during each invocation of
{\algmsme}, we start from scratch by ignoring any mass that the jobs
may have accumulated in the previous rounds. We now analyze the
performance of {\algsuuiObl}.

If $t \ge 2\topt$, with one invocation of {\algmsme} using $t$, let
$x$ be the number of jobs that get at least $1/96$ mass. The total sum
of masses over the jobs is at most $x\cdot 1 + (n-x)\cdot 1/96$
because the mass that any job accumulates is at most $1$. From
Theorem~\ref{thm:exist}, we know that there exists an {\oblivs} of
length $t$, with a total sum of mass at least $n/16$.  Now according
to Lemma~\ref{lem:msme}, {\algmsme} has an approximation ratio of
$1/3$.  Thus,
\[
x\cdot 1 + (n-x)\cdot 1/96 \ge 1/3 \cdot n/16.
\]
It follows that $x\ge n/95$. Since each invocation of \\{\algmsme} makes
at least $1/95$ of the jobs accumulate $1/96$ mass, it is sufficient
to invoke {\algmsme} at most $66\log n$ times until all jobs
accumulate at least $1/96$ mass.

To prove that {\algsuuiObl} terminates in polynomial time, we first
bound the value of $\topt$. Let $p_{min} = \min_{i,j} p_{ij}$.
Obviously, if we let the jobs accumulate sufficient mass one by one by
assigning all machines to a single job at any step, then every job
accumulates a mass of at least $1$ within a time interval of
$\ceil{\frac{n}{p_{min}}}$. This implies that $\topt =
O(\frac{n}{p_{min}}\log n)$. Since $t$ is doubling every iteration in
{\algsuuiObl}, $O(\log n+ \log \frac{1}{p_{min}})$ different $t$ values
will be ``probed'' before the algorithm terminates. With each $t$
value, we invoke {\algmsme} at most $66\log n$ times, and each such
invocation runs in polynomial time. We conclude that algorithm
{\algsuuiObl} terminates within time polynomial in the size of the input.
We have thus proved:
\begin{lemma} \label{lem:constant} For Problem {\suui}, one can
  compute in polynomial time an {\oblivs} of length $O(\log n)\topt$
  in which every job accumulates a mass of at least $1/96$.
\end{lemma}

\begin{theorem} \label{thm:sch.ind} For Problem {\suui}, within
  polynomial time, we can compute an {\oblivs} whose expected makespan
  is within a factor of $O(\log^2 n)$ of the optimal.
\end{theorem}
\begin{proof}
  Using Lemma~\ref{lem:constant}, we first compute an {\oblivs}
  $\Sigma_o$ of length $T = O(\log^2 n)\cdot\topt$ in which every job
  accumulates a mass of at least $1/96$. The infinite repetition of
  $\Sigma_o$, $\Sigma_o^{\infty}$, is the {\oblivs} we want. Treating
  the execution of $\Sigma_o^{\infty}$ during each step interval of
  $[k\cdot T+1, (k+1)\cdot T]$, where $k=0,1,\ldots$, as one
  iteration, by Proposition~\ref{pro:bounds} we know that every job
  has a success probability of at least $\frac{1}{24e}$ during each
  iteration. Within $O(\log n)$ iterations, all jobs are finished with
  high probability. Thus, the expected makespan of $\Sigma_o^{\infty}$
  is within $O(\log^2 n)$ of $\topt$. We now formalize this argument.

  Let random variable $X$ be the iteration number when all jobs are
  finished. We bound the expected value of $X$ below.
  \begin{eqnarray*}
    E[X] & = & \sum_{i=0}^{\infty} \Pr(X > i) \\
    & = & \sum_{i=0}^{362\log n -1} \Pr(X > i) + \sum_{i=362\log n}^{\infty} \Pr(X > i) \\
    & \le & 362\log n\cdot 1 + \sum_{i=362\log n}^{\infty} n\cdot(1-\frac{1}{96e})^{i} \\
    & = & 362\log n+ n\cdot(1-\frac{1}{96e})^{362\log n}\cdot\sum_{i=0}^{\infty} (1-\frac{1}{96e})^{i} \\
    & \le & 362\log n + \frac{96e}{n},
  \end{eqnarray*}
  where the third inequality follows because every job has a
  probability $\frac{1}{96e}$ of success within each iteration, and the
  last inequality follows by summing the geometric series and the fact
  that $(1-\frac{1}{96e})^{181} < 1/2$. This completes the proof of the
  theorem.
\end{proof}
\fi

\newcommand{\aspc}{{\sf AccuMass-C}}

\section{Jobs with precedence \\constraints} \label{sec:sch.precede}
In this section, we study {\suu} when there are non-trivial precedence
constraints on the jobs. We first present in \S\ref{sec:sch.chain} a
polylogarithmic approximation algorithm for the case when the
constraints form disjoint {\em chains}, and then extend the results in
\S\ref{sec:sch.forest} to 
the more general case when the constraints form directed forests.  All
of the schedules we compute are oblivious.

\subsection{Disjoint chains}
\label{sec:sch.chain}
We consider {\suu} in the special case where the dependency graph $C$
for the jobs is a collection of disjoint chains $C=\{C_1,\cdots,
C_l\}$.  We refer to this problem as {\suuc}.  If job $j_1$ precedes
$j_2$ according to the constraints, we write $j_1 \prec j_2$.

At a high level, our approach to solve {\suuc} is to first compute an
{\oblivs} of near-optimal length in which every job has a constant
probability of successful completion, then {\em replicate} this
schedule sufficiently many times to conclude that all the jobs are
finished with high probability within a desired makespan bound.  We
first consider the problem of accumulating a constant success
probability for each job.  As in the independent jobs case, we will
use the notion of mass instead of the actual probability.  However, we
need to take into account the dependencies among the jobs.  Therefore,
we formulate the following problem {\aspc}: Given the input for
{\suuc}, compute an {\oblivs} with minimum length $T$, subject to two
conditions: (i) Every job $j$ accumulates a mass of at least $1/2$
within $T$; (ii) If $j_1 \prec j_2$, $j_1$ must already accumulate
mass $1/2$ before any machine can be assigned to $j_2$.  Condition
(ii) captures the intuition that if $j_1$ has a low probability of
successful completion before step $t$, then the probability that $j_2$
is eligible for execution at step $t$ would be small; so it does not
make much sense to assign machines to $j_2$ prior to $t$ in the
{\oblivs}.

\junk{ The reason for Condition
  (ii) will be clear when we later use such a schedule to finish the
  jobs.  }

The following is a relaxed linear program (LP1) for \\{\aspc}.  Let
$x_{ij}$ denote the number of steps during which machine $i$ are
assigned to $j$. Let $d_j$ be the number of steps during which there
is some machine assigned to $j$.
\begin{eqnarray}
  \mbox{(\bf LP1)}\qquad\quad \min &t & \nonumber\\
  \mbox{s.t.} \quad \sum_{i\in M} p_{ij}x_{ij} &\ge &1/2 \quad \forall j\in J \label{eq:req}\\
  \sum_{j\in J} x_{ij} & \le & t \quad \forall i\in M \label{eq:load}\\ 
  \sum_{j\in C_k}  d_j &\le & t \quad C_k \in C \label{eq:dilation}\\
  0 \le x_{ij} & \le & d_j \quad \forall i,j \label{eq:proc}\\
  d_j &\ge & 1 \quad \forall j \label{eq:comp}
\end{eqnarray}
Some comments on (LP1) are in order. Equation~\ref{eq:req} enforces
Condition (i). Equation~\ref{eq:load} bounds the {\em load} on every
machine, which we define below. Equation~\ref{eq:dilation} bounds the
time length on each chain constraint. Finally Equation~\ref{eq:proc}
ensures that each job accumulates its mass during the $d_j$ steps when
there is some machine assigned to it. Let $T^*$ be the optimal value
for (LP1) above.

Note that in (LP1) we do not have any condition to prevent two
different jobs from two precedence chains to be scheduled on the same
machine at the same step. We use the term {\em \pseus} to capture such
``schedules'', in which different jobs from different precedence
chains may be scheduled to the same machine simultaneously.

\begin{definition}
  A {\bf \pseus} of {\bf length} $T\in \mathbb{Z^+}$ $\cup\, \infty$ is a
  collection of assignment functions, $\{f_{t}: M\to 2^J\,|\, 1\le t<
  T+1\}$.
\end{definition}

Hence, an assignment function of a {\pseus} may map a machine to a set
of jobs. In this sense, a {\pseus} may not be feasible; we address
this issue later when describe how to transform a {\pseus} to an
appropriate {\oblivs}. An {\oblivs} is a {\pseus} in which the value
of $f_{t}$ is a single element.

\begin{definition}
  Given a {\pseus} $\Sigma_g$ of (finite) length $T$, $\{f_{t}: M\to
  2^J\,|\, 1\le t< T+1\}$, the {\bf load of a machine} $i$ is defined
  as the total number of times that a job is scheduled on $i$ in
  $\Sigma_g$. Formally, the load of machine $i$ is $\sum_{1\le t< T+1}
  |f_t(i)|$. The {\bf load} of $\Sigma_g$ is defined as the maximum
  load of any machine.
\end{definition}
We remark that a {\pseus} of length $T$ may have a load greater than
$T$.

\begin{theorem} \label{thm:round} Within polynomial time one can round
  an optimal feasible solution to (LP1), and obtain a {\pseus} for
  Problem {\aspc} whose length and load are both $O(\log m)T^*$.
\end{theorem}
\begin{proof}
  Obviously (LP1) is feasible because one can assign machines to each
  job for a finite steps so that the job can accumulate a mass of
  $1/2$.  Let $\{x_{ij}, d_j, t\}$ be one {\em optimal} solution to
  (LP1). (Note that $t$ is equal to $T^*$.) Our efforts mainly concern
  the rounding procedure, i.e., obtaining a feasible {\em integral}
  solution from the fractional solution without blowing up $t$ too
  much. We then describe how to get a {\pseus} from an integral
  solution to (LP1). We differentiate between two cases.

The first case is when $t \ge |J|=n$. We round each $x_{ij}$ and $d_j$
up by setting $x_{ij}^* = \ceil{x_{ij}}$ and $d_j^* = \ceil{d_j}$. We
obtain a feasible integral solution with approximation factor 2 since
we have
\begin{eqnarray*}
  \sum_{i\in M} p_{ij}x^*_{ij} & \ge & 1/2 \quad \forall j\in J,\\
  \sum_{j\in J} x^*_{ij} &\le& t + n \le 2t \quad \forall i\in M,\\
  \sum_{j\in C_k}  d^*_j &\le& t+n \le 2t \quad C_k \in C,\\
  x^*_{ij} &\le& d^*_j \quad \forall i,j. 
\end{eqnarray*}
The second case is when $t <|J|=n$. We make use of some results from
network flow theory for our rounding in this case. Notice that
although we target for a mass of $1/2$, any constant smaller than
$1/2$ will do as well because we can always scale every variable up to
reach that target, sacrificing only a constant factor. In our
presentation below, we use many such scale-up operations.  (We haven't
tried to optimize the constants.)  For a given job $j$, if $\sum_{i\in
M, x_{ij} \ge 1} p_{ij}x_{ij} \ge 1/4$, we can round these $x_{ij}$'s
to the next larger integer. Since $\ceil{x_{ij}} \le 2x_{ij}$, this
only incurs a factor of 2 blow up in $t$. Thus, we only need to
consider those jobs $j$ such that $\sum_{i\in M, x_{ij} \ge 1}
p_{ij}x_{ij} \le 1/4$, which implies that $\sum_{i\in M, x_{ij} < 1}
p_{ij}x_{ij} \ge 1/4$. Observe that $\sum_{i\in M, p_{ij} <
\frac{1}{8m}, x_{ij} < 1} p_{ij}x_{ij} < 1/8$,
which implies\\ $\sum_{i\in M, p_{ij} \ge \frac{1}{8m}, x_{ij} < 1}
p_{ij}x_{ij} \ge 1/8$.

We bucket these $p_{ij}$'s into at most $B=\ceil{\log (8m)}$ intervals
$(2^{-(k+1)}, 2^{-k}]$ ($k=0,1,\ldots$). For a bucket\\ $b: (2^{-(b+1)},
2^{-b}]$, if $\sum_{p_{ij} \in {\mbox{\scriptsize bucket }}b} x_{ij} <
1/32$, we remove this bucket from further consideration. Note that the
sum of $p_{ij}x_{ij}$ over all removed buckets is at most
$1/16$. Hence for the $p_{ij}$'s in the remaining buckets, we still
have\\ $\sum_{i\in M, p_{ij} \ge \frac{1}{8m}, x_{ij} < 1} p_{ij}x_{ij}
\ge 1/16$.

For each job $j$, there is a bucket $b_j: (2^{-(b_j+1)}, 2^{-b_j}]$
such that $\sum_{p_{ij} \in {\mbox{\scriptsize bucket }}b_j} x_{ij}\ge
\frac{2^{b_j}}{16B}$.  
Denote the sum on the left side of the above inequality by $D_j$. If
necessary, we scale all the $x_{ij}$'s (and other variables) up by a
factor of 32, so that all $D_j \ge 1$.  We then round $D_j$ down to
$\floor{D_j}$. These operations only cost us a constant factor in
terms of approximation. Thus for the ease of the presentation below,
we assume that the $D_j$'s are integral and let $D = \sum_{j\in J}
D_j$.

We now construct a {\em network-flow} instance as follows (see
Figure~\ref{fig:flow}). We have one node for each job $j$, one node
for each machine $i$, a source node $u$, and a destination node $v$.
We add an edge $(i,j)$ for each $x_{ij}$ contributing to the
computation of $D_j$'s. We orient the edge $(i,j)$ from $j$ to $i$, with
edge capacity $\ceil{d_j}$.\junk{ (causing a blow-up factor of at most
$2$). (Wlog, $d_j$ is integral. Since if $d_j < 1$, round it to
  1; if $d_j \ge 1$, round it to next larger integer.  Doing this only
  blow up the dilation by at most a factor of 2.)} From each machine
node $i$, add an edge toward $v$, with capacity $\ceil{2t}$.\junk{ (2 is
added due to the rounding of $d_j$).} For each job node $j$, add an
edge from $u$ to $j$, with capacity $D_j$.  \junk{The demand on $u$ is
  $D$.}

\begin{figure}[h]
\centering
\includegraphics[width=3.3in]{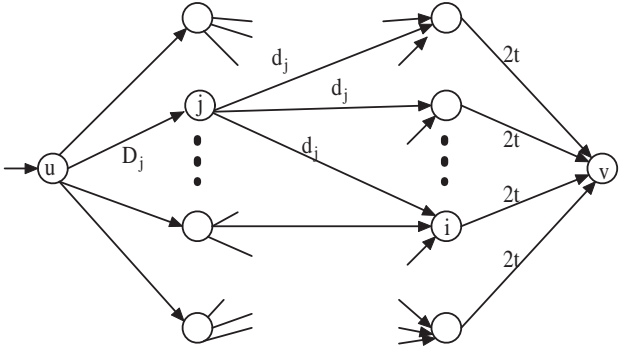}
\caption{A network flow instance for the rounding of an optimal
  solution to (LP1)}
\label{fig:flow}
\end{figure}

The argument before the construction shows that  a flow of demand
$D$ at $u$ can be pushed through the network, where the $x_{ij}$'s
specify such a feasible flow. $D$ is actually the maximum flow of the
network (consider the cut where one side consists of $u$ alone). From
Ford-Fulkerson's theorem~\cite{FordFulkerson62,Cormen-algorithms}, we
know that there exists an {\em integral} feasible flow when the
parameters are integral, as in our instance. We take such an integral
flow value on edge $(j,i)$ as our rounded solution $x^*_{ij}$.
Furthermore, the integral solution obtained observes the following
identities.
\junk{
\begin{eqnarray*}
 & \sum_{i\in M} p_{ij}x^*_{ij} \ge\frac{1}{16\ceil{\log(8m)}} \quad
  \forall j\in J, \quad & \sum_{j\in J} x^*_{ij} \le \ceil{2t}  \quad \forall i\in M\\
 & \sum_{j\in C_k}  \ceil{d_j} \le  \ceil{2t}  \quad C_k \in C, \quad &
  x^*_{ij} \le \ceil{d_j} \quad \forall i,j.
\end{eqnarray*}
}
\begin{eqnarray*}
 \sum_{i\in M} p_{ij}x^*_{ij} &\ge&\frac{1}{16\ceil{\log(8m)}} \quad
  \forall j\in J, \\
 \sum_{j\in J} x^*_{ij} &\le& \ceil{2t}  \quad \forall i\in M,\\
 \sum_{j\in C_k}  \ceil{d_j} &\le&  \ceil{2t}  \quad C_k \in C,\\
  x^*_{ij} &\le& \ceil{d_j} \quad \forall i,j.
\end{eqnarray*}
Raising all the values by a factor of $O(\log m)$, we obtain an
integral feasible solution $\{\hat{x}_{ij}, \hat{d_j}, \hat{t}\}$,
where $\hat{t} = O(\log m)T^*$. \junk{we get a relaxed schedule whose
  length and load are bounded by $O(\log m)T^*$.}

We now describe how to construct from the integral solution a {\pseus}
$\Sigma_s$ whose length and load are both bounded by $\hat{t} = O(\log
m)T^*$.  Consider a job $j$ in a chain $C_k \in C$. Given the
$\hat{x}_{ij}$'s, let $L_j = \max_i \hat{x}_{ij}$.  Let $\psi_j =
\sum_{j_0:j_0 \prec j} L_{j_0}$. We assign the machines to $j$ within
a step interval of length $L_j$ from step $\psi_j+1$ to $\psi_j +
L_j$, using each machine $i$ $\hat{x}_{ij}$ times. In other words, the
assignment functions for chain $C_k$ are specified as follows. For any
job $j$ and machine $i$, if $\hat{x}_{ij} > 0$, $f^k_t(i) = \{j\}$ for
$t \in [\psi_j+1, \psi_j+\hat{x}_{ij}]$. This can be done because each
machine is assigned to $j$ at most $L_j$ times and different machines
can be assigned to $j$ at the same step.  After we define the
$f^k_t(\cdot)$ for every chain $C_k \in C$, we define the assignment
functions for $\Sigma_s$ as
\[
f_t(i) = \cup_{k:
  C_k\in C} f^k_t(i) \quad \mbox{for } i\in M, t \in [1, \hat{t}].
\]
Recall that the range of the assignment functions for a {\pseus} is a
set of jobs. This completes the proof of the theorem.
\end{proof}

We now relate {\aspc} to {\suuc}.  Recall that $T^*$ is the optimal
value of (LP1) we write for Problem {\aspc}, and $\topt$ is the
expected makespan of an optimum schedule $\Sigma$ for Problem
{\suuc}. We now bound the value $T^*$ in terms of $\topt$ in
Lemma~\ref{lem:link}.  This lemma, together with
Theorem~\ref{thm:round} immediately yields a {\pseus} that solves
{\aspc} with load and length within $O(\log n)$ factor of $\topt$.

\begin{lemma} \label{lem:link}
  $T^* \le 16\topt$. \qed
\end{lemma}

\begin{proof}
  The following linear program is the same as (LP1), except that $1/2$
  is replaced by $1/16$ and $t$ is replaced by $2\topt$. We argue that
  this linear program is feasible.
\begin{eqnarray*}
  \sum_{i\in M} p_{ij}x_{ij} &\ge &1/16 \quad \forall j\in J \\
  \sum_{j\in J} x_{ij} & \le & 2\topt \quad \forall i\in M\\
  \sum_{j\in C_k}  d_j &\le & 2\topt \quad C_k \in C \\
  x_{ij} & \le & d_j \quad \forall i,j \\
  d_j &\ge & 1 \quad \forall j \\
  x_{ij} &\ge&  0 \quad \forall i,j 
\end{eqnarray*}

Consider the first $2\topt$ execution steps using an optimal schedule
$\Sigma$. Let random variable $X_{ij}$ be the number of steps in which
$i$ is assigned to $j$. Let random variable $Y_j$ be the total number
of steps when there is some machine assigned to $j$.  We know from
Theorem~\ref{thm:sim} that with probability at least $1/4$, $j$
accumulates at least $1/4$ mass within $2\topt$ steps. This amounts to
the fact that the expected accumulated mass for $j$ is at least
$1/16$. Thus
\[
\sum_{i\in M} p_{ij}\cdot E[X_{ij}] \ge 1/16.
\]

Since in $\Sigma$ a machine is assigned to at most a job at any 
step, $\sum_{j\in J} X_{ij} \le 2\topt$. So
\[
\sum_{j\in J} E[X_{ij}] \le 2\topt.
\]

Since we are considering only $2\topt$ steps of $\Sigma$, we have
$\sum_{j\in C_k} Y_j \le 2\topt$. Obviously, $X_{ij} \le Y_j$. Taking
the expectation, we have
$$\sum_{j\in C_k} E[Y_j] \le 2\topt$$ and
$$E[X_{ij}] \le E[Y_j].$$

We conclude that $x_{ij} = E[X_{ij}]$ for $i\in M, j\in J$ and $d_j =
E[Y_j]$ for $j\in J$ form a solution to the linear program. Raising
this solution by a factor of $8$, we obtain a solution to (LP1).  This
means that a $t$ of value $16\topt$ is achievable in (LP1). We have
thus proved that $T^* \le 16\topt$.  This completes the proof of the
lemma.
\end{proof}

\begin{theorem} \label{thm:chain.exist} A {\pseus} with length and
  load bounded by $O(\log m)\cdot\topt$ can be computed within
  polynomial time, such that: (i) Every job $j$ accumulates at least
  $1/2$ mass. (ii) If $j_1 \prec j_2$, $j_2$ can only begin the
  accumulation after $j_1$ accumulates $1/2$ mass. \qed
\end{theorem}
\junk{
\begin{proof} 
  This follows from Lemma~\ref{lem:link} and Theorem~\ref{thm:round}.
\end{proof}
}
In the remainder of this section, we describe how to convert a
{\pseus} obtained from Theorem~\ref{thm:chain.exist} to a feasible
schedule.  According to Theorem~\ref{thm:chain.exist}, we can compute
a {\pseus} $\Sigma_s$ of length $O(\log m)\cdot\topt$ in which every
job accumulates a mass of at least $1/2$, and hence a success
probability of at least $\frac{1}{2e}$. Moreover, if $j_1 \prec j_2$,
no machine is assigned to $j_2$ until $j_1$ has accumulated $1/2$ such
mass.  We now convert $\Sigma_s$ to a (feasible) oblivious schedule
$\Sigma_o$ in two steps. \junk{ We describe these two steps briefly and
refer the reader to the appendix for details.}  \junk{, without
lengthening the schedule too much.}
\begin{enumerate}
\item We use the elegant random delay technique of~\cite{LeightonMR94,ShmSteWei94} to delay
the start step of the execution for each chain appropriately and
  obtain a new {\pseus} $\Sigma_{s,1}$ in which the number of jobs
  scheduled on any machine at any step is $O(\frac{\log
  (n+m)}{\log\log (n+m)})$.  The randomized schedule can also be
  derandomized using techniques
  from~\cite{Raghavan88a,SchSieSri95,ShmSteWei94}. We then ``flatten''
  $\Sigma_{s,1}$ to obtain an {\oblivs} $\Sigma_{o,1}$, sacrificing a
  factor of $O(\frac{\log (n+m)}{\log\log (n+m)})$ in the schedule's
  length.
\item To obtain the final {\oblivs} $\Sigma_o$, we take the {\oblivs} $\Sigma_{o,1}$ from above and replicate each
  step's machine assignment $O(\log n)$ times, so that all jobs will
  be finished with high probability.
\end{enumerate}

We now describe in detail the two steps that convert a {\pseus} to a
feasible {\oblivs}.  Since the second step is simpler, we describe it
first.

\noindent{\bf Schedule replication:} \junk{If $\Sigma_s$ is already
  {\em feasible}, in the sense that no more than one job is scheduled
  on the same machine at the same step, then we can simply replicate
  $\Sigma_s$} We first replicate $\Sigma_{o,1}$ at each step by a
factor of $\sigma= 16\log n$ to get another {\oblivs} $\Sigma_{o,2}$.
More precisely, let $T$ denote $\Sigma_{o,1}$'s length and let
$g_t(\cdot)$'s be the assignment functions of $\Sigma_{o,1}$. We
define the assignment functions $f_t(\cdot)$'s of $\Sigma_{o,2}$ as
follows. For any $t\in [1, \sigma\cdot T]$, $f_t(\cdot) =
g_\tau(\cdot)$, where $\tau = \floor{\frac{t-1}{\sigma}}+1$. Note that
if $\Sigma_{o,1}$ can be specified in space polynomial in the size of
the input, as we will show in the ``delay'' step, so can
$\Sigma_{o,2}$.

We define yet another {\oblivs} $\Sigma_{o,3}$ of length $n$ as
follows. Topologically sort the jobs according to the precedence
constraints, e.g., appending the precedence chains one after another,
and let $j_1,\ldots, j_n$ be the jobs in the sorted order. The
assignment functions $h_t(\cdot)$'s for $\Sigma_{o,3}$ are specified
as follows.  $\forall i\in M, h_t(i) = j_t$, where $1\le t\le n$. Now
the final {\oblivs} we want is $\Sigma_o =
\concat{\Sigma_{o,2}}{\Sigma^{\infty}_{o,3}}$. In other words,
{\oblivs} $\Sigma_o$ is simply the replicated $\Sigma_{o,1}$ followed
by assigning all the machines to some job at each step.

\junk{
The specification of
$f_t(\cdot)$ for $t> \sigma\cdot T$ is less formal, which will be
provided shortly, after we discuss the expected makespan of
$\Sigma_o$, whose length is $\sigma T = O(\log n\log m)\topt$.  
}

We now analyze the expected makespan of $\Sigma_o$. If all jobs are
successfully completed within step $\sigma T$, the expected makespan
is at most $\sigma T$. The probability that this does not happen is at
most $n(1-\frac{1}{2e})^{\sigma} < 1/n^2$. Notice also that from step
$\sigma T+1$ on, $\Sigma_o$ assigns all the machines to a single job
at each step periodically (due to $\Sigma_{o,3}$, with a period length
of $n$). The expected number of steps for a job to be completed is at
most $\topt$ if all the machines are assigned to it. Since we
periodically assign the machines to any fixed job, on average, it
takes at most $(n\topt)$ steps to complete any fixed job.  Hence, on
average, it takes at most $n^2\topt$ steps to complete all the jobs
using the assignment functions beyond step $\sigma T$. The expected
makespan of $\Sigma_o$ is thus at most
\[
(1-1/n^2) \sigma\cdot T + 1/n^2\cdot (\sigma\cdot T+ n^2 \topt).
\]
As we will prove shortly, $T = O(\log m \frac{\log (n+m)}{\log\log
  (n+m)})\cdot\topt$ and $\sigma=16\log n$. We conclude that the
expected makespan of $\Sigma_o$ is $O(\log n\log m \frac{\log
  (n+m)}{\log\log (n+m)})\cdot\topt$.

\smallskip

\noindent{\bf Converting {\pseus} $\Sigma_s$ to an {\oblivs}:} We now
address the issue when the computed {\pseus} $\Sigma_s$ from
Theorem~\ref{thm:chain.exist} is not yet feasible, that is, when some
machine is assigned to more than one job at the same step. We claim
that we can convert $\Sigma_s$ to an {\oblivs} $\Sigma_{o,1}$ by
sacrificing a factor of $O(\frac{\log (n+m)}{\log\log (n+m)})$.

Let $\Pi_{max}$ be the load of $\Sigma_s$, i.e., the maximum number of
jobs assigned to any machine. A result by Shmoys, Stein and Wein on
job shop scheduling problem~\cite[Lemma 2.1]{ShmSteWei94} states that
if we {\em delay} the starting step of each chain by an integral
amount independently and uniformly chosen from $[0,\Pi_{max}]$, the
resulting {\pseus} has no more than $O(\frac{\log (n+m)}{\log\log
  (n+m)})$ jobs scheduled on any machine during any step. We now
explain what we mean by the term {\em delay}. Recall that in the last
paragraph of the proof for Theorem~\ref{thm:round}, we first specify a
function $f_t^k$ for each constraint chain $C_k \in C$, and then
define assignment function for $\Sigma_s$ as $f_t = \cup_{k} f^k_t$.
Suppose that a chain $C_k$ is delayed by an amount of $\phi_k$, the
assignment function $g_t^k$ for chain $C_k$ is modified as follows.
$\forall i\in M$, if $t\le \phi_k, g_t^k(i) = \emptyset$; otherwise,
$g_t^k(i) = f_{t-\phi_k}^k(i)$. And the assignment function for the
schedule is defined as $f_t = \cup_{k} g^k_t$. To make our
presentation self-contained, we now outline the argument for the bound
of $O(\frac{\log (n+m)}{\log\log (n+m)})$ below.

Fix a step $t$ and a machine $i$. Let $p = \Pr[${at least $\tau$ units
  of processing are scheduled on machine $i$ at step $t$}$]$.  Note
that a job $j$ could be scheduled in multiple steps, and each job is
unit-step, it is equivalent to say that there are multiple processing
units of job $j$. There are at most ${\Pi_{max} \choose \tau}$ ways to
choose those $\tau$ processing units. Focus on a particular choice of
$\tau$ units. If these units are from different chains, the
probability that they are all scheduled at step $t$ is at most
$(\frac{1}{\Pi_{max}})^\tau$ since we choose the delay independently
and uniformly from $[0, \Pi_{max}]$. Otherwise, the probability is $0$
because our {\pseus} can never assign two units from the same chain to
the same machine at the same step.  Therefore,
\begin{eqnarray*}
  p &\le& {\Pi_{max}\choose \tau} \left(\frac{1}{\Pi_{max}}\right)^\tau \\
  & \le & \left(\frac{e\Pi_{max}}{\tau}\right)^\tau \left(\frac{1}{\Pi_{max}}\right)^\tau \\
  & \le & \left(\frac{e}{\tau}\right)^\tau
\end{eqnarray*}

If $\tau = \alpha\frac{\log (n+m)}{\log\log (n+m)}$, then $p <
(n+m)^{-(\alpha-1)}$.  Let $L_{max}$ be the length of the longest
chain according to $\Sigma_s$. The probability that {\em any} machine
at {\em any} step is assigned at least $\alpha\frac{\log
  (n+m)}{\log\log (n+m)}$ jobs is bounded by
$m(\Pi_{max}+L_{max})(n+m)^{-(\alpha-1)}$.  With the assumption, which
we will remove shortly, that $\topt$ is bounded by a polynomial in
$(n+m)$, $\Pi_{max}+L_{max}$ is bounded by a polynomial in $(n+m)$ as
well. If we choose $\alpha$ to be sufficiently large, then with high
probability, no more than $\alpha\frac{\log (n+m)}{\log\log (n+m)}$
jobs are scheduled on any machine at any step.

Shmoys, Stein and Wein~\cite{ShmSteWei94} also derandomize the
algorithm so that $O(\log (n+m))$ jobs can be scheduled on any machine
simultaneously, based on results
by~\cite{RagTho85,RagTho87,Raghavan88a}. Schmdit, Siegel and
Srinivasan~\cite{SchSieSri95} give a different derandomization
strategy and obtain a collision bound matching the randomized
algorithm, i.e., $O(\frac{\log (n+m)}{\log\log (n+m)})$ machines
simultaneously for any machine. We denote this (derandomized) {\pseus}
by $\Sigma_{s,1}$, whose length is at most twice that of
$\Sigma_s$. According to Theorem~\ref{thm:chain.exist}, $\Sigma_s$'s
length is $O(\log m)\cdot\topt$, it follows that we can ``flatten''
$\Sigma_{s,1}$ out to obtain an {\oblivs} $\Sigma_{o,1}$ whose length
is $O(\log m\frac{\log (n+m)}{\log\log (n+m)})\cdot\topt$, in which
each machine is assigned to one job at any step. We comment that the
{\em random delay} technique originates in~\cite{LeightonMR94} when
they study the job shop scheduling problem.

\smallskip

\noindent {\bf Reducing $\topt$:} We now address the issue that
$\topt$ is not always bounded by a polynomial in $(n+m)$. We make use
of a trick from \cite[Section 3.1]{ShmSteWei94}. Consider the {\pseus}
$\Sigma_s$ computed in Theorem~\ref{thm:chain.exist}. For each job
$j$, let $l_{ij}$ be the number of steps in which machine $i$ is
assigned to $j$ and $L_j$ be $\max_i l_{ij}$. Denote $\max_j L_j$ by
$L$. We know that all machines are assigned to $j$ within a window of
length $L_j$.\junk{ Let $L_{max}$ be the length of the longest chain
  according to the relaxed schedule, and $\Pi_{max}$ be the maximum
  load on a machine.}  Let $\beta = nm$. Round each $l_{ij}$ down to
the nearest multiple of $\frac{L}{\beta}$, and denote this value by
$l'_{ij}$. We therefore can treat the $l'_{ij}$ as integers in
$\{0,\ldots, \beta\}$. A schedule for this new problem can be
trivially rescaled to one with the real values $l'_{ij}$.  Since
$\beta = nm$, the schedule now {\em effectively} has a length (and
load) bounded by a polynomial in $(n+m)$. Hence our discussions of the
random delay and derandomization hold now. Let $\Sigma'$ be the
resulting feasible {\oblivs}, with length bounded by $O(\log
m\frac{\log (n+m)}{\log\log (n+m)})\topt$ and load bounded by $O(\log
m)\topt$.  To get a feasible {\oblivs} $\Sigma_{o,1}$ so that every
job accumulates $1/2$ mass, we {\em insert} $(l_{ij}-l'_{ij})$ units
of processing to $\Sigma'$. The insertion can be done in a way that
preserves the precedence constraints, i.e., if $j_1\prec j_2$, then no
machine can be assigned to $j_2$ before $j_1$ accumulates $1/2$ mass.
Since each insertion lengthens $\Sigma'$ by an amount $\le
\frac{L}{nm}$ and we have at most $nm$ such insertions, the length of
the schedule is increased by at most $L$. The loads on the machines
are the same as before the rounding. Note that $L$ is bounded by
$\Pi_{max}$, which is $O(\log m)\topt$. We thus have obtained a
feasible {\oblivs} $\Sigma_{o,1}$ whose length is $O(\log m\frac{\log
(n+m)}{\log\log (n+m)})\topt$, in which every job accumulates a {\em
constant} mass.  Finally, we use the {\em replication} technique
discussed earlier in this section to obtain the desired schedule.

\begin{theorem} \label{thm:chain} For Problem {\suuc}, there exists a
  poly-nomial-time algorithm to compute an {\oblivs} schedule with
  expected makespan within a factor of \\$O(\log m\log n\frac{\log
  (n+m)}{\log\log (n+m)})$ of the optimal. \qed
\end{theorem}

For independent jobs, i.e., when the constraints $C$ in Problem
{\suuc} is empty, we can prove a bound for oblivious schedules that
slightly improves over the result stated at the end of
\S\ref{sec:sch.ind}.

\begin{theorem} \label{thm:sch.ind.imp} For Problem {\suui}, there exists a
  poly-nomial-time algorithm to compute an {\oblivs} schedule with
  expected makespan within a factor of\\ $O(\log n\cdot\log(\min\{n,
  m\}))$ of the optimal. \qed
\end{theorem}

\begin{proof}
  Let (LP2) be the linear program obtained from (LP1) by removing
  constraints \ref{eq:dilation}, \ref{eq:proc}, \ref{eq:comp}, and
  $T_2^*$ be (LP2)'s optimal value.  We first show that one can round
  an optimal feasible solution to (LP2), and obtain an {\oblivs} for
  Problem {\aspc}, whose length, and hence load, are both
  $O(\log(\min\{n,m\}))\cdot T_2^*$. 

  For Problem {\suui}, Condition (ii) of {\aspc} is void. We thus
  don't need constraints \ref{eq:dilation}, \ref{eq:proc},
  \ref{eq:comp} when writing the linear program. The rounding in the
  proof of Theorem~\ref{thm:round} gives an $O(\log m)$ blow-up. If
  $m\ge n$, we can do a better analysis for the rounding procedure.
  Since there are $n+m$ non-trivial constraints in (LP2), there are at
  most $n+m$ nonzero values in any basic feasible
  solution~\cite{Bertsimas97,Schrijver86}. In an optimal solution
  $\{x_{ij}, t\}$ (which is basic feasible), we may assume without
  loss of generality that for any machine $i$, there exists a $j$ such
  that $x_{ij} > 0$.  Otherwise, we may remove that machine from
  consideration in (LP2).  From here, we conclude that the number of
  machines $i$ that have at least two $x_{ij} >0$ is at most $n$.
  When we round $x_{ij}$'s, we only need to consider these machines
  $i$ with at least two $x_{ij} >0$. Then the same rounding procedure
  in the proof of Theorem~\ref{thm:round} gives a factor $O(\log n)$
  blow-up because for each job, we only need to consider $O(\log n)$
  buckets.

  We conclude that one can obtain an integral feasible solution
  $\{\hat{x}_{ij}, \hat{t}\}$ where $\hat{t} =
  O(\log(\min\{n,m\}))\cdot T_2^*$. Furthermore, from $\{\hat{x}_{ij},
  \hat{t}\}$, one can construct a (feasible) {\oblivs} for Problem
  {\aspc}, whose length, and hence load, are $\hat{t} =
  O(\log(\min\{n,m\}))\cdot T_2^*$. This is because the load on each
  machine is bounded by $\hat{t}$ according to Equation~\ref{eq:load}
  and the jobs are independent. Hence the machine assignment can be
  done in such a way that no more than one job is scheduled on any
  machine at any step.

We thus have an {\oblivs} in which every job accumulates a {\em
constant} mass within time that is at most $O(\log(\min\{n,m\})$ times
optimal.  We now apply the schedule replication step and obtain the
desired bound.
\end{proof}

\subsection{Tree-like precedence constraints} \label{sec:sch.forest}
Our algorithm for tree-like precedence constraints uses techniques
from~\cite{KumarEtAl05}, who extend the work of~\cite{ShmSteWei94} on
scheduling unrelated parallel machines with chain precedence
constraints to the case where there are tree-like precedence
constraints by decomposing the directed forests into $O(\log n)$
collection of chains. To state their result, we first introduce some
notations used in~\cite{KumarEtAl05}.  Given a dag $G(V,E)$, let
$d_{in}(u)$ and $d_{out}(u)$ denote the in-degree and out-degree,
respectively, of $u$ in $G$. A {\em chain decomposition} of $G$ is a
partition of its vertex set into subsets $B_1, \ldots, B_{\lambda}$
(called blocks) such that: (i) The subgraph induced by each block
$B_i$ is a collection of vertex-disjoint directed chains; (ii) For any
$u,v\in V$, let $u\in B_i$ be an ancestor of $v\in B_j$. Then, either
$i<j$, or $i=j$ and $u$ and $v$ belong to the same directed chain of
$B_i$; (iii) If $d_{out}(u) > 1$, then none of $u$'s out-neighbors are
in the same blocks as $u$. The {\em chain-width} of a dag is the
minimum value $\lambda$ such that there is a chain decomposition of
the dag into $\lambda$ blocks. We now state the decomposition result.

\begin{lemma}[\cite{KumarEtAl05}, Lemma 1]\label{lem:decompose}
  Every dag whose underlying undirected graph is a forest has a chain
  decomposition of width $\gamma$, where $\gamma\le 2(\ceil{\log
    n}+1)$. The decomposition can be computed within polynomial time.
\end{lemma}

\junk{, where each block is a collection
of chains and the precedence constraints are preserved appropriately.}

Using Lemma~\ref{lem:decompose}, we simply decompose a given directed
forest into at most $\gamma= O(\log n)$ blocks, and within each block,
apply our algorithm for the chain case
(Theorem~\ref{thm:chain}). Since the optimal expected makespan on any
subgraph (subset of jobs) is a lower bound for that of the whole graph
(whole set of jobs), this approach gives up another factor of $\log
n$. We have thus obtained
\begin{theorem} \label{thm:forest} For Problem {\suu}, if the
  dependency graph $C$ is a directed forest, there exists a
  polynomial-time algorithm to compute an {\oblivs} schedule with
  expected makespan within a factor of $O(\log m\log^2 n\frac{\log
  (n+m)}{\log\log (n+m)})$ of the optimal.
\end{theorem}

\junk{, and obtain an $O(\log m\frac{\log^3
  n}{\log\log n})$ approximation algorithm.  However, as
in~\cite{KumarEtAl05}, we believe a more careful argument will allow
us to preserve the same bound as in the chain case.
}

When the precedence constraints form a collection of {\em out trees}
(rooted trees with edges directed away from the root) or {\em in
trees} (defined analogously), we can obtain an improved approximation
algorithm by again following the ideas of~\cite{KumarEtAl05}.  More
specifically, we decompose the out/in trees into $O(\log n)$ blocks;
then randomly delay each chain by an amount of steps chosen uniformly
from $[0, O(\Pi_{max}/\log n)]$ (this step can be derandomized in
polynomial time); and prove that with high probability, at most
$O(\log n)$ jobs can be scheduled on any machine simultaneously.  

\begin{theorem} \label{thm:tree} For Problem {\suu}, if the dependency
  graph $C$ is a collection of out/in trees, there exists a
  polynomial-time algorithm to compute an {\oblivs} schedule with
  expected makespan within a factor of $O(\log m\log^2 n)$ of the
  optimal.
\end{theorem}

\section{Open problems} \label{sec:sch.open} In this paper, we have
presented polylogarithmic approximation algorithms for the problem of
multiprocessor scheduling under uncertainty, for special classes of
dependency graphs.  We believe that our bounds are not tight; in
particular, we conjecture that a more careful analysis will improve
the approximation ratios by an $O(\log n)$ factor in each case.  It
will also be interesting to obtain approximations for more general
classes of dependencies, and to consider online versions of our
scheduling problem.

\bibliographystyle{abbrv}
\bibliography{schedule}
\end{document}